\documentclass[letterpaper]{article} 
\usepackage{aaai2026}  
\usepackage{times}  
\usepackage{helvet}  
\usepackage{courier}  
\usepackage[hyphens]{url}  
\usepackage{graphicx} 
\usepackage{natbib}  
\usepackage{caption} 
\usepackage{algorithm}
\usepackage{algorithmic}

\usepackage{newfloat}
\usepackage{listings}
\DeclareCaptionStyle{ruled}{labelfont=normalfont,labelsep=colon,strut=off} 
\floatstyle{ruled}
\newfloat{listing}{tb}{lst}{}
\floatname{listing}{Listing}
\title{AdFL: In-Browser Federated Learning for Online Advertisement}
\author {
    Ahmad Alemari\thanks{Ahmad is also with 
    the College of Engineering and Computer Science,
    Jazan University, Jazan, Saudi Arabia}, 
    Pritam Sen, 
    Cristian Borcea
}
\affiliations {
    New Jersey Institute of Technology\\
    Newark, New Jersey, USA\\
    \{aaa376, ps37, borcea\}@njit.edu
    
}

\newcommand{\modaa}[1]{#1}
\newcommand{\newaa}[1]{#1}
\newcommand{\newaar}[1]{#1}

\usepackage{xcolor}

\usepackage{subcaption}

\usepackage{booktabs}
\usepackage{makecell}
\usepackage{amssymb}

\begin{document}

\maketitle

\begin{abstract}

Since most countries are coming up with online privacy regulations, such as GDPR in the EU, online publishers need to find a balance between revenue from targeted advertisement and user privacy. One way to be able to still show targeted ads, based on user personal and behavioral information, is to employ Federated Learning (FL), which performs distributed learning across users without sharing user raw data with other stakeholders in the publishing ecosystem. This paper presents AdFL, an FL framework that works in the browsers to learn user ad preferences. These preferences are aggregated in a global FL model, which is then used in the browsers to show more relevant ads to users. AdFL can work with any model that uses features available in the browser such as ad viewability, ad click-through, user dwell time on pages, and page content. The AdFL server runs at the publisher and coordinates the learning process for the users who browse pages on the publisher's website. The AdFL prototype does not require the client to install any software, as it is built utilizing standard APIs available on most modern browsers. We built a proof-of-concept model for ad viewability prediction that runs on top of AdFL. We tested AdFL and the model with two non-overlapping datasets from a website with 40K visitors per day. The experiments demonstrate AdFL's feasibility to capture the training information in the browser in a few milliseconds, show that the ad viewability prediction achieves up to 92.59\% AUC, \newaar{and indicate that utilizing differential privacy (DP) to safeguard local model parameters yields adequate performance, with only modest declines in comparison to the non-DP variant}.

\end{abstract}

\section{Introduction}

In the online publishing ecosystem, revenue from online ads incentivizes publishers to enable data collection about users and their activities, which is then shared with ad agencies to deliver targeted ads. While this data-sharing practice maximizes the revenue for publishers, it raises significant user privacy concerns, prompting regulations such as GDPR~\cite{gdpr2016} and CCPA~\cite{ccpa2018} that limit data collection and sharing without explicit user consent~\cite{smith_study_2024}. When users decline consent for data collection and sharing, advertisers are not able to show targeted ads. Instead, they can show only contextual ads, which result in lower revenue for publishers. Federated Learning (FL) could be used by publishers to find a balance between targeted ads and revenue.

\modaa{
Existing FL systems for online ads face integration challenges with current ad servers. For instance, ~\cite{wu_fedctr_2022} suggests a model using two hierarchical servers, but it does not fit seamlessly with existing systems such as Google Ads Manager, which require a key-value bid structure instead of embeddings. Implementing a publisher ad server increases costs and complexity due to necessary communication with external ad servers. The conversion model in ~\cite{wei_fedads_2023} relies on vertical FL and requires conversion metrics extending browser data to extensive external data, which complicates deployment. 
~\cite{wu_fedctr_2022,bian_feynman_2023,zhao_privacy-aware_2020} depend on synthetic data instead of real-world publisher data. Some systems ~\cite{hartmann_federated_2019,brennaf_comparative_2023,alzamel_webflex_2024, lian_webfed_2022,ting_yang_feasibility_2022} support in-browser training, but they are not tailored for online ads and lack data collection capabilities.
}

This paper proposes AdFL, an FL framework that works in the browser to balance user privacy and publisher's revenue from online ads.
AdFL avoids the complexity of hierarchical servers~\cite{wu_fedctr_2022} by using a single server at the publisher to aggregate a global model and not requiring an ad server. AdFL runs asynchronously in the browser and does not require additional mechanism to integrate with the existing ad ecosystem. Furthermore, it provides learning based on ad metrics data available in the browser, such as ad viewability, ad click-through, user attention on the ads, user dwell time on pages, page content, ad data and meta-data, etc. AdFL is based on horizontal FL which is more feasible to be deployed in conjunction with real-world applications than vertical FL because global models are built without modifying the model architecture, as data samples are distributed among users.

AdFL can work with any model that uses features available in the browser and with any aggregation algorithm. The publisher chooses the model and hosts the FL aggregation server. If other publishers want to use the same model, the same server can be used for all of them. However, different models require separate servers. To keep the local models in sync with the global model and minimizes the delay for heterogeneous users, AdFL uses first-party cookies to track global model versions at the client side. AdFL is designed to work without requiring the client to install any software, as it leverages common APIs available in most browsers. 

\newaa{
AdFL directly addresses challenges specific to browser-based FL, such as operating within constrained environments, ensuring seamless data collection and preprocessing within the browser, and managing computational and memory overheads on diverse devices, while considering ad-serving requirements. Unlike other studies that simply load the FL model and data into the browser, AdFL is a comprehensive FL pipeline designed to operate within the browser. The pipeline starts with data collection, focusing on user interactions, ads, and page-related features. Ensuring data quality and performing effective preprocessing are crucial steps in this pipeline. AdFL is designed to work efficiently in real-time in resource-constrained environments (i.e., it uses about 50MB of memory and performs inference in a few ms). AdFL is a flexible framework that enables model designers to adapt their workflows to support different models, ad scripts, page resources, browsers, and devices. 

}

We implemented an AdFL prototype in the browser that consists of five components, interacting asynchronously with online ads. The components are responsible for gathering the browsing sessions and data, tracking online ads, processing data, and  managing the model. 
The implementation uses Tensorflow.js
and JavaScript for the browser components. The in-browser JavaScripts utilize standard browser APIs, such as MutationObserver API, to ensure AdFL is compatible with all major browsers. Furthermore, given the high efficiency and asynchronous nature of the third-party ad code, the prototype was implemented to capture the required data in real-time. We built several versions of a proof-of-concept ad viewability model to run on top of AdFL and tested them with two non-overlapping datasets of 10-days and 30-days from a website with 40K visitors per day. The models are capable of handling a larger set of input features on more capable devices by utilizing a concatenation layer that dynamically activates inputs based on the available data features. The application of these models is to estimate the ad relevance for a given user on a given webpage, based on
IAB's viewability definition~\footnote{\url{https://www.iabuk.com/news-article/quick-qa-viewability}}. The application dynamically adjusts the duration of the ad display to show ads that are more relevant for users and thus increase publisher revenue. 

Our experiments demonstrate the AdFL's feasibility in practice and its ability to capture ads' data and process them in the browser,
while maintaining the dataset in the browser to be utilized for ad measurement models. 
AdFL is capable to process data and run an inference in less than 3ms for desktop browsers and less than 7ms for mobile browsers. 
In addition, the viewability model on top of AdFL achieves a very good AUC of up to 92.59\% in FL settings. 
\newaar{Moreover, AdFL can employ differential privacy (DP) to protect local model parameters, and the models work well, with only modest reductions compared to the non-DP versions.}

To summarize, the key novel contributions are:
\begin{itemize}
    \item The design of a novel FL framework, AdFL, that enables in-browser FL model training and inference with seamless local data collection and preprocessing for online ads.
    \item An ad viewability prediction model, working on top of AdFL, that can be leveraged to increase the viewability of the ads relevant for users, and thus maximize the publisher's revenue.
    \item A prototype implementation of AdFL that we demonstrated to be able to collect ad data in real-time and train effective FL models for privacy-preserving ad-targeting applications. 
\end{itemize}

The paper is organized as follows. Section~\ref{sec:related work} presents background on the online ad ecosystem and related work. Section~\ref{sec: system} describes the AdFL architecture. Section~\ref{sec: data and model} explains the browser-collected data and the model used to test AdFL. Section~\ref{sec:implementation} details the AdFL client implementation. Section~\ref{sec: experiment} reports system and model evaluation results. Section~\ref{sec: conclusion} concludes the paper.

\section{Background and Related Work} \label{sec:related work}
This section provides background information for ad loading mechanisms involving header bidding, followed by a discussion of relevant literature.

\begin{table*}[t]
\centering
\small
\setlength{\tabcolsep}{2pt}
\renewcommand{\arraystretch}{1.0}
\begin{tabular*}{\textwidth}{@{\extracolsep{\fill}} lcccccccc}
\toprule
\textbf{Approach} 
  & In-browser
  & \makecell{No Adv./\\DSP coop.}
  & \makecell{Browser-only\\data}
  & \makecell{Ad-specific\\Integration}
  & \makecell{Low\\latency}
  & \makecell{Live\\ads}
  & \makecell{Easy\\integ.}
  & Scalable \\
\midrule
Vertical FL\\\cite{wei_fedads_2023,bian_feynman_2023}       & \textbullet & \texttimes & \texttimes & \texttimes & \textbullet & \textbullet & \texttimes & \textbullet \\ \hline
Privacy-aware FL~\cite{zhao_privacy-aware_2020}            & \textbullet & \checkmark & \texttimes & \texttimes & \texttimes & \texttimes & \textbullet & \textbullet \\ \hline
FedCTR~\cite{wu_fedctr_2022}                              & \textbullet & \checkmark & \textbullet & \textbullet & \textbullet & \texttimes & \textbullet & \textbullet \\ \hline
Content Recommender~\cite{tan_federated_2020}             & \texttimes & \checkmark & \texttimes & \texttimes & \checkmark & \texttimes & \checkmark & \checkmark \\ \hline
Centralized Ad Models\\\cite{kalra_reserve_2023,wang_viewability_2015} & \texttimes & \textbullet & \texttimes & \checkmark & \texttimes & \checkmark & \checkmark & \checkmark \\ \hline
Generic Browser FL\\\cite{lian_webfed_2022,hartmann_federated_2019}   & \checkmark & \checkmark & \texttimes & \texttimes & \textbullet & \texttimes & \textbullet & \checkmark \\ \hline
\textbf{AdFL}                                              & \checkmark & \checkmark & \checkmark & \checkmark & \checkmark & \checkmark & \checkmark & \checkmark \\
\bottomrule
(\checkmark = supported; \textbullet = partial; \texttimes = none)
\end{tabular*}
\caption{Comparison between AdFL and related work}
\label{tab:comparsion-relatedwork}
\end{table*}

\subsection{Background}\label{subsec:background}

This subsection explains how online ads are delivered to publishers. We also discuss the publisher's limited role in ad selection or improvement.

In decentralized ad auctions, such as header bidding~\cite{kalra_reserve_2019}, as soon as the user loads a page that has online ads enabled, the header bidding script is loaded to the browser and starts broadcasting the ad placements to the auction participants. Within a specific time frame, the script gathers the bids in a key-value structure and sends this bids' structure to the ad server to select the winning ad based on its own priorities. The ad server delivers the winning ad to the publisher's webpage. 
Throughout this procedure, the publisher has limited influence over the different stages of the auction. Its control is mainly restricted to providing a predefined list of direct ads to the ad server, which is how the open-sourced ads auction~\footnote{Such as Prebid: \url{ https://docs.prebid.org/overview/intro.html}, Apache 2.0 License} adapts to the ad server~\cite{mackenzie_longest_2023}. 

The structure of the winner ad includes the encapsulation of monitoring and tracking scripts. The innermost part is the advertiser loading mechanism that loads the ad creative, which is typically hosted on a server or a CDN. Once the ad is loaded to the browser, it is placed inside a secured iframe, and the advertiser script runs to load the ad creative. Both the advertisers and the ad server have monitoring and tracking scripts to ensure the integrity and validity of ad delivery, and to detect fraud. Additionally, an essential part is the ad activation, which acknowledges the winning bid to the advertisers to secure the transaction and downloads the ad creative. 

This ad structure limits the access of publishers to online ads within the browser, a restriction rigorously enforced by policies like cross-origin constraints within a secure iframe. While AdFL efficiently captures the ad creative in real-time, it does not disclose the ad's content. Consequently, AdFL abstains from targeting the content of Ads, thereby maintaining the ads within the browser environment, conserving device resources by preventing the extraction of links to ad content, and adhering to the constraints of the online ad market by refraining from activating ad content links.

\subsection{Related Work}
This subsection discusses FL for online ads, centralized models for online ads, and FL within browser, highlighting the differences between these works and our approach.

The study in~\cite{wei_fedads_2023} introduces a vertical FL model utilizing actual data from both an online publisher and an advertising agency to learn the conversion rate metrics. 
This approach necessitates the cooperation of advertisers for model training and inference because the input data features are distributed among publishers and other ad-related stakeholders such as demand-side platforms (DSPs).
In contrast, AdFL employs horizontal FL, where data records are distributed across users, thereby eliminating the need for external data and collaboration with other entities.

A similar work~\cite{bian_feynman_2023} proposes a model enhancing the conversion rate for mobile users. Apart from the conversion rate metrics that require data not available in the browser, this model requires the cooperation of the advertisers to provide the additional data. AdFL, on the other hand, focuses on data that is available in the browser and avoids the complexity of collaboration with advertisers. 

The work in~\cite{zhao_privacy-aware_2020} presents a privacy model trained on users' data from those who opt in to share their data. Then, this model is served using FL to the users opting out of sharing their data. AdFL does not require any user to share their data, and we assume that all users opt out of sharing their data.

The paper~\cite{wu_fedctr_2022} proposes an FL model to improve click-through rate (CTR) prediction across multiple platforms, including the browser, by aggregating user behaviors. The model generates user embeddings from local behaviors on each platform, which are then aggregated on a user server and shared with the ad server for CTR prediction.
AdFL is designed to work with many models, whereas the solution above is specific to CTR prediction. In addition, AdFL enables FL models whose local training runs in the users' browsers, and the aggregation server works at the publisher. The AdFL client serves the outcomes of the model to the user's browser.
In~\cite{tan_federated_2020}, the authors proposed a content-based FL recommendation system for general online content, 
which cannot be applied directly to online ads in the browser as AdFL.

Several works proposed centralized models for online ads.
The works in~\cite{kalra_reserve_2023,kalra_reserve_2019} focuse on enhancing the ad reward mechanism for publishers, while the ones in ~\cite{wang_webpage_2019,wang_viewability_2015} propose centralized models for viewability prediction of online ads. Furthermore, ~\cite{xu_ukd_2022} focus on enhancing centralized models for conversion rate estimation. These models are not privacy-preserving solutions, whereas AdFL solves this issue by using FL. These models could be redesigned to work in an FL fashion, and then they could run on top of the AdFL framework. 

The existing work on FL within browsers~\cite{lian_webfed_2022,alzamel_webflex_2024,brennaf_comparative_2023,chen_openfed_2023,ting_yang_feasibility_2022,hartmann_federated_2019} has significant limitations, especially concerning their data collection and preprocessing capabilities. Furthermore, none of these works focuses on real-time data collection and model inference for online ads.
The study in~\cite{hartmann_federated_2019} examines FL to rank browser histories. The proposed method is computationally expensive, making it impractical in our settings where the publisher-site implementations are restricted by the user's visit duration.
The studies presented in~\cite{lian_webfed_2022,chen_openfed_2023,ting_yang_feasibility_2022} introduce frameworks that enable the implementation of FL directly within web browsers. These frameworks are different from AdFL because they do not have ad-specific support and do not collect data in the browser. They are tested through simulations/emulations with standard image classification datasets.
The study in~\cite{alzamel_webflex_2024} employs a peer-to-peer structure, which is not practical in the existing online publishing ecosystem. 
Furthermore, the research in~\cite{brennaf_comparative_2023} assesses the efficacy of FL across various web browsers, uncovering important insights regarding FL's compatibility with these platforms. In comparison, AdFL is oriented toward a real-world application of FL within the realm of online advertising.

\newaar{
Table~\ref{tab:comparsion-relatedwork} provides a synthesis of the related work by selecting common comparative features that are applicable across the most relevant studies covered in this section.
}

\section{System Design} \label{sec: system}

\begin{figure}[tbp]
    \centerline 
    {\includegraphics[width=0.95\linewidth]{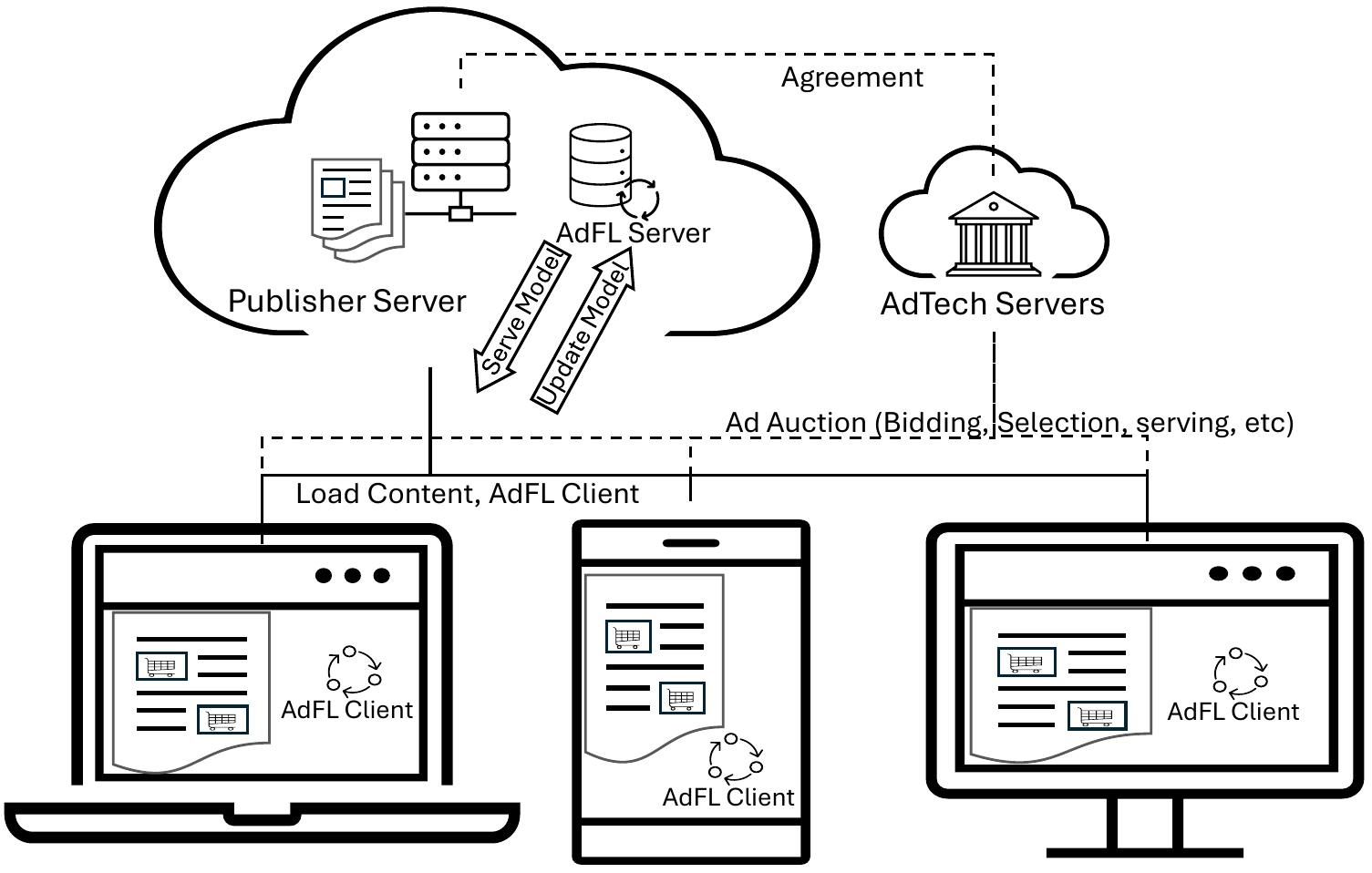}}
    \caption{AdFL Overview}
    \label{fig:sys_overview}
\end{figure}

This section presents the design of AdFL and its threat model. As shown in Figure~\ref{fig:sys_overview}, we consider a typical online publishing ecosystem, where the publisher server has agreements with AdTech servers to deliver ads to clients. AdFL enables client devices to collect contextual and ad-specific data for each ad, creating one data sample per ad. Since this data may disclose private information about the user browsing behavior and preferences, we adopt the FL paradigm, allowing clients to use their dataset to train models locally and upload the model parameters to the server.

The AdFL server operates either at the publisher or as a third-party server enabled by the publisher. The client-side software, called AdFL Client, runs scripts downloaded from the AdFL server in the browser following AdFL requirements. Each AdFL Client operates independently to collect and store data, using the data to train a local model, which is then uploaded to the AdFL server. The AdFL server aggregates these local models into a generalized global model, which is subsequently distributed to all AdFL clients.

In AdFL, we address a real-world scenario where the publishers seek to increase revenues by aligning the ads effectively with user interests. This typically involves transmitting user data externally, which raises privacy concerns. Instead, AdFL enables the publisher to analyze user interactions directly within the browser using an end-to-end FL system, allowing control over ad behavior based on FL model predictions. As long as the prediction works well, the advertisers get to target the users, and the users get more relevant ads. In the rare case when mispredictions happen, the users will see less relevant ads.

To manage the complexity of this process, we propose several novel components in AdFL, as no existing FL system fully supports this scenario.
Additionally, for certain components, we customize existing solutions to meet specific needs, providing a comprehensive end-to-end system design and implementation. The main novelties of AdFL are summarized as follows:

\begin{itemize}
\item AdFL runs a script to capture ad-specific web elements in the browser, allowing the client to collect ad-specific data.
\item AdFL allows the clients to manage session history by collecting and processing data specific to each session.
\item AdFL allows the publisher to adjust ad behavior by tracking multiple metrics in the browser using the client-side software.
\item AdFL enables developers to configure the system to enhance model utility by adjusting parameters according to the training environment.
\item AdFL preprocesses diverse datasets in the browser, ensuring consistent feature representation across all clients.
\item AdFL uses optimization techniques to minimize the recollection of invariant data, thereby reducing computational and storage overhead.
\end{itemize}

\subsection{Threat Model}\label{subsec:threat_model}
\newaar{
In our system, there are two key entities:
\begin{itemize}
    \item Data Owners (DO): End-users whose browsers execute scripts for the AdFL Client. Each DO trains and runs inferences using private session data, then sends model updates to the AdFL server. Their main privacy concern is preventing the leakage of sensitive user behavior and interaction patterns.
    
    \item Model Owner (MO): Typically, the publisher or a trusted third party manages the AdFL server. The MO's main priorities are preserving the aggregated model's integrity and ensuring accurate inference results. We assume MO to follow the AdFL protocol, acting honest-but-curious.
    
\end{itemize}
Our threat model $TM_{AdFL}$ assumes secure communication between DOs and MO via HTTPS. Each DO retains its own session and data locally and cannot access other DOs' data. The MO, while correctly executing the FL protocol, is an honest-but-curious adversary analyzing updates through various attacks or altering aggregation to infer individual behaviors. 
FL secures the raw data, ensuring it never exits the DO's computer. Although MO might infer information from local model updates, this risk can be mitigated by incorporating differential privacy ~\cite{yang_dynamic_2023} alongside FL, a method that has shown to be effective with negligible impact on performance.
}

\subsection{AdFL Server}\label{subsec:adfl_server}
The AdFL Server is hosted at the publisher and interacts with clients visiting the publisher's webpages. The server has two components: the Server Manager and the Aggregator. The Server Manager is a web service that performs two key functions: fulfilling client requests for global model parameters and handling incoming updates from clients with new local model parameters. The Aggregator combines the model parameters received from the clients during each training round. The Server Manager and the clients use a tag to identify the latest global model version. Depending on the model tag received in a client request, the Server Manager provides the latest version of the global model parameters, if a newer version is available. Otherwise, the client request waits for the Aggregator to release a new global model update. 
The clients receive the latest global model and train it using their local data for a predetermined number of rounds before they are allowed to send model updates.
By default, AdFL employs FedAvg
for aggregation, but it can accommodate other aggregation techniques. \modaa{In addition, existing solutions (e.g., differential privacy ~\cite{yang_dynamic_2023,qi_towards_2024}) could be integrated in the system to protect against gradient-based attacks. }

The AdFL server hosts the AdFL client scripts that are accessible to the users' browsers for download. The home page of the publisher's website (and potentially other pages) contains a script that initially communicates with the AdFL server to download the required scripts for in-browser computation at the client side.

\subsection{AdFL Client}\label{subsec:adf_client}
\begin{figure}[tbp]
    \centering
    {\includegraphics[width=1\linewidth]{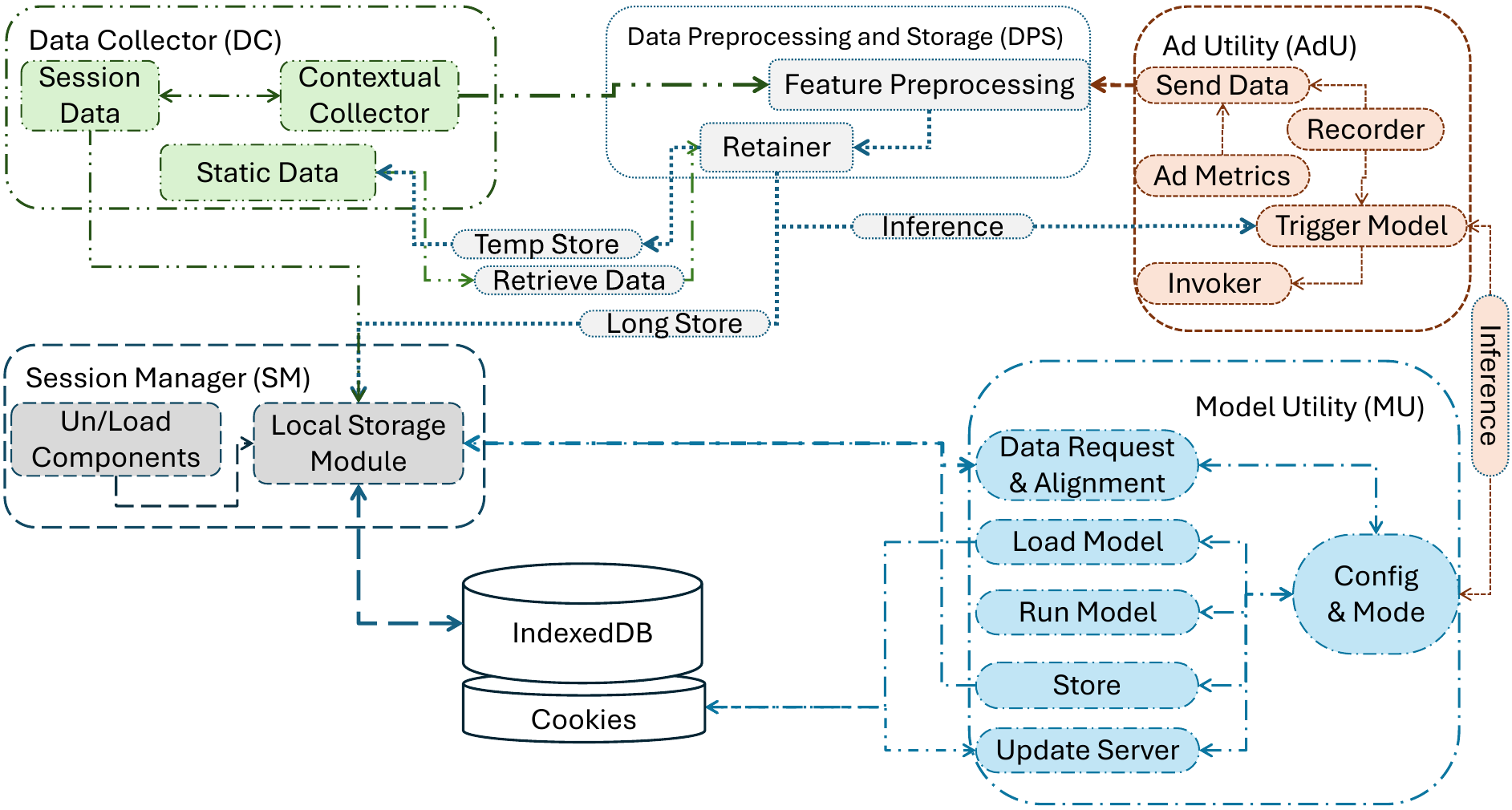}}
    \caption{In-browser AdFL Client Design}
    \label{fig:prototype_components}
\end{figure}

Since AdFL is designed to work in the browser, most of the complexity is in the client software. The AdFL client comprises five essential components, as depicted in Figure~\ref{fig:prototype_components}: the Session Manager (SM), the Data Collector (DC), the Ad Utility (AdU), the Data Preprocessing and Storage (DPS), and the Model Utility (MU). These components are implemented as scripts and are retrieved from the server. Upon the initial page request of a user session, the SM script is executed first, subsequently downloading the remaining components, which persist in the browser as long as the user remains on the publisher's page. The SM initializes the storage and invokes the DC, AdU, and MU. The DC is executed to collect contextual data on each page the user navigates to on the publisher's site. With the loading of an ad on a page, the AdU is executed to capture the ads data. Both DC and AdU send their data to DPS for data processing and subsequently send the output to SM for storage. The MU oversees the model and its configuration settings and begins the training process according to its configuration. It is also activated by AdU for model inference.

The components of the AdFL client are systematically divided into modules, ensuring that data collection exhibits distinct recurrence intervals for contextual data compared to ad data. Contextual data is updated upon page requests, whereas ad data undergoes multiple refreshes within a single page view. Given that the data collected is in its raw form and necessitates preprocessing tailored for specific subsets of the entire dataset, the DPS encompasses diverse data processing techniques and compiles each dataset record appropriately. Executing the model involves several subsequent tasks and specific configurations; hence, the MU consolidates all model execution-related tasks into a singular unit. In the following, we detail each of these components.

\subsubsection{Session Manager (SM)} \label{subsec:smc}

\modaa{
SM manages AdFL client sessions, each comprising multiple user visits to the same publisher's pages within a set timeframe. It oversees loading/unloading of AdFL components and browser storage. On the first page request of a session, SM loads script-based AdFL components from the server. AdFL stores data samples, session details, configurations, and model parameters in browser storage. SM deploys the database schema, handling all read/write operations and ensures all data is saved before unloading a webpage.
}

\subsubsection{Data Collection (DC)} \label{subsec:dcc}

\modaa{
DC captures browser data, including page, user, and session details, storing them in raw and preprocessed forms for efficient processing. It has three modules: contextual data collector, static data module, and session data aggregator. The contextual data collector gathers user and page information such as user agent, cookies, browser type, OS, URL, and page content. This data is sent to DPS for preprocessing, and stored by the static data module, which stores unchanging data for ads to avoid repetitive collection, as user and page data usually remain static unless the user changes their browser or system. 

Session data characterization is limited to the active session, causing incomplete user-page pattern representation. The session data aggregator addresses this by integrating features from current and past sessions. The DC must store raw records of previous sessions in browser storage, including page requests and timestamps. When a new session begins, these records are retrieved, user-page patterns are calculated and preprocessed, then the updated data is temporarily stored before being saved for long-term use.
}

\subsubsection{Ad Utility (AdU)}\label{subsec:aduc}

AdU collects ad-specific features and metrics to prepare data samples for model training and inference. Given the time-sensitive nature of online ads, AdU is designed to efficiently capture this data immediately as the ads load. 

\modaa{
AdU records data from online ads on visited pages, capturing refreshed ad instances. It tracks metrics like ad viewability and compiles these into data features for each ad sample. Additionally, AdU triggers the MU to perform inference on the ad sample and awaits results. The invoker then uses these results to perform publisher-defined actions, such as modifying the ad's refresh interval causing the current ad to expire sooner than scheduled or extend beyond schedule.
}

The recorder module utilizes standard browser APIs, particularly event listeners, to gather data and metadata related to online ads. Event listeners monitor specified events on designated target HTML elements, activating a predefined function when such an event occurs. When an ad loads, the event listener for that ad placement is triggered, capturing the necessary data and initiating the metrics module to establish its event listeners for that ad instance. These metrics are event listeners to monitor interactions with the ad. For example, the ad metrics listeners are activated when the ads meet the conditions for viewability metrics. This event trigger confirms positive labels for the metrics 
and then communicates the metrics label and ad identifier to DPS through the send data module. Both the recorder and metrics modules transmit their respective data independently for processing.

To distinguish the categories and purposes of ad data transmitted from AdU for data preprocessing request, the module responsible for sending data attaches particular flags to the request. This enables the DPS to interpret the data as belonging to one of three categories: storage, update, or inference. The storage flag indicates that a new data sample should be saved to the browser storage after preprocessing. The update flag signifies that the data contains new metrics related to a previously transmitted data sample, identified by a unique ad identifier. The inference flag indicates that the data is intended for model inference, requiring a preprocessed copy to be sent back to the trigger model module while retaining another copy for storage.

\newaa{
The trigger model is responsible for initiating inference for each loaded ad. For every ad instance, 1) it retrieves the preprocessed data sample from the DPS component, 2) sends the data sample to the MU by triggering an inference request, and 3) receives the prediction from the model, passing it to the invoker.
For example, the predictions from the ad viewability model are used to adjust the ad refresh rate. The refresh rate, typically set to 30 seconds, dictates how long an ad remains displayed before a new instance is loaded. If the model predicts an ad as non-viewable, the refresh rate is adjusted, allowing for more relevant ads to be shown to users. This dynamic adjustment improves the ad mechanism, beyond just the impression, and user engagement, ultimately leading to increased revenue for the publisher.

}

\subsubsection{Data Preprocessing and Storage (DPS)}\label{subsec:preprocessing}

DPS enables the preprocessing of raw data collected within the browser, typically sourced from DC and AdU. Furthermore, DPS supports the compilation and preservation of dataset records. DPS is composed of two submodules: feature preprocessing and retainer. The feature preprocessing is tasked with processing the raw data received from DC and AdU, ensuring that the data is properly prepared for training. The retainer stores the processed data within the browser by employing the SM's local storage module.

DPS uses a predefined list of features, allowing the model designer to regulate them. This list includes each feature's name, preprocessing methodologies, and attributes such as maximum, minimum, or default values necessary for specific preprocessing techniques. For instance, the feature preprocessing module uses MinMaxScaling to transform the page height, using given attributes by the model designer. Meanwhile, hashing is applied to categorical string values, like the browser's user agent, independent of the feature's attributes and dataset values.

\modaa{
DPS uses the retainer module to maintain processed data in the browser by assembling data samples. Each sample contains lists of values for ad features, including contextual and ad data with metric labels. These samples are sent to SM for storage and a copy is returned to AdU for inference if flagged. The retainer module obtains processed ad data from AdU and contextual data from DC to complete the feature set. Initially, ad metric labels are set to negative, indicating unmet conditions. When metric updates arrive, the retainer forwards the information to SM to update browser storage. 
}

\subsubsection{Model Utility (MU)} \label{subsec:muc}

\modaa{
MU serves model training and inference, 
allowing model designers to load and deploy models with predefined configurations. Through these configurations, model designers establish data dynamics, model architecture, and device conditions to optimize training and inference processes. Data factors include sample quantity, distribution, label count, temporal relevance, and sampling methods. Designers also have control over model dimensions, training intensity, early stopping, model history, and versioning. Device performance is optimized by considering battery, memory, and computational power, ensuring high-quality training and inference.

When the configuration conditions are met, the MU starts the training process by obtaining data from SM, and split the data typically to training, validation, and testing subsets 
After loading the latest global model from AdFL server and using the model version kept in cookies, MU performs the training for the given number of rounds. The MU then saves the model in the browser and sends the local model parameters and version tag to the server.
}

\section{Data and Model Design}\label{sec: data and model}

To demonstrate AdFL, we collected user, page, and ad data from a real publisher website and built a proof-of-concept model executed on top of AdFL. This section presents the dataset attributes, preprocessed features used as model input, and the model architecture.

\subsection{Data features}\label{subsec:data features}

The data attributes are generally independent of the publisher and are classified into four categories: user, page, session, and ads. Table~\ref{tab:features_pii_categories} provides a summary of these data attributes and outlines the categories of Personally Identifiable Information (PII) contained within each category. AdFL is capable of utilizing these PII data attributes for local model training, without leaving the browser.

The session data, which encapsulates the users' page visit history, is not directly captured but derived from accumulated metrics such as the number of pages visited and the duration since the last visit. AdFL computes this session data for each data sample in relation to prior visits, see Section~\ref{subsec:dcc} for more. The user-related data comprises identifiers such as browser versions, user agents, operating system, and IP addresses. The content data encompasses elements such page titles, content, and URL.

AdFL uses event-driven triggers and the browser API to track ad loading and refreshing. Ads are generally hosted externally and retrieved via iframes, with cross-origin policies limiting content visibility. While AdFL captures ad creatives in real-time, manual feature engineering is needed to extract insights, identify the ad tech agency, and understand metrics. A key element in ad creatives is the activation link, which loads the ad content and alerts the ad tech agency of a successful bid.

\begin{table}[t!]
    \setlength{\tabcolsep}{0.3mm}
    \renewcommand{\arraystretch}{0.95}
    \centering
    \begin{tabular}{l|c|l}
        \hline
        \textbf{Category} & \textbf{\# Features} & \textbf{PII Categories (\#)} \\ 
        \toprule
        User    & 22 & Identifiers (6), Location (3) \\ 
        Page    & 16 & Personal Attributes (3) \\ 
        Session & 28 & Identifiers (3), Behavioral (4) \\ 
        Ad      & 33 & Identifiers (5) \\ \hline
    \end{tabular}
    \caption{Model Features and PII Categories}
    \label{tab:features_pii_categories}
\end{table}

\subsection{Model Design} \label{subsec:model_design}

We designed a lightweight proof-of-concept model, engineered to execute inference in real-time during the ad loading phase within the browser environment. The primary aim of this model is to forecast the duration of ad viewability following deployment. Typically, an ad lasts thirty seconds; however, AdFL's invoker, as noted in \ref{subsec:muc}, can adjust this by changing the interval refresh rates, which could boost publisher revenue by optimizing ad display time and {triggering load of} ads that are of higher interest to the user. Similarly, other models and use cases can be built using the data captured by AdFL, which is general for most online ads.

\begin{figure}[tbp]
    \centering
    \includegraphics[width=0.95\linewidth]{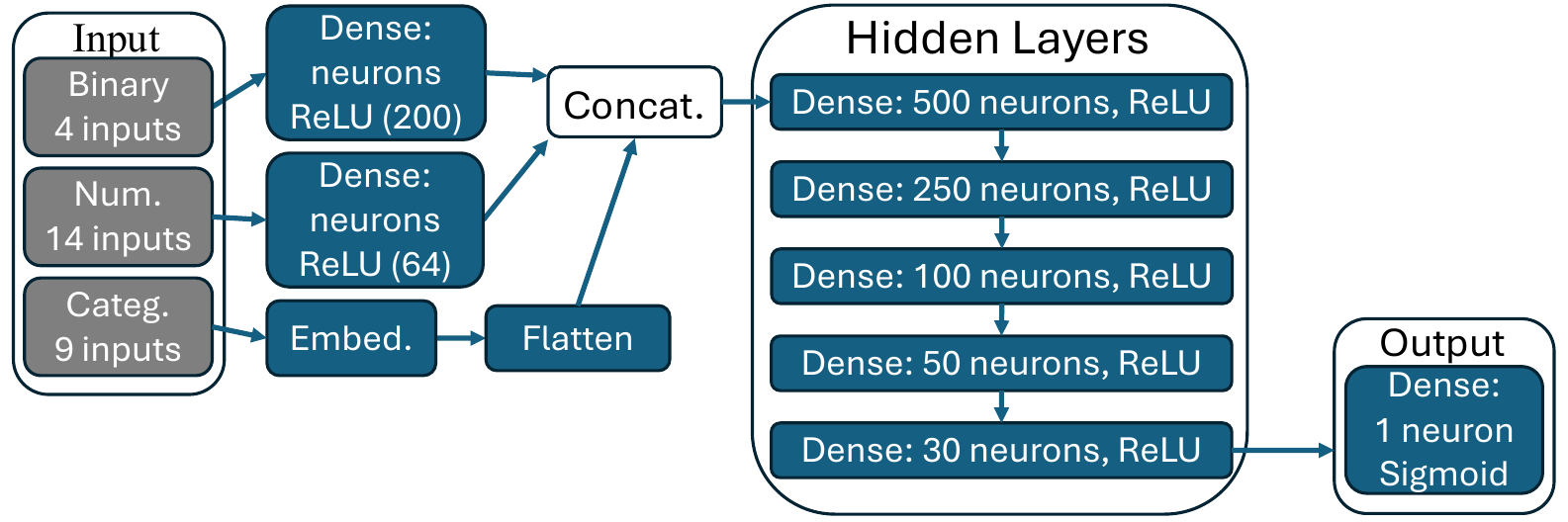}
    \caption{Model Design}
    \label{fig:model-mini}
\end{figure}

The invoker is organized according to input types, comprising numerical, binary, and categorical data, as illustrated in Fig.~\ref{fig:model-mini}.  The input is partitioned according to data type. Given the feature processing supported by AdFL, we apply hash encoding to categorical features and MinMAX scaling to numerical ones. Consequently, our model receives 3 binary, 14 numerical, and 9 categorical features as inputs.
The model architecture uses input layers that connect to dense layers for both numerical and binary features, while categorical features pass through embedding layers followed by flatten layers. These processed inputs are then merged using concatenation layers. After this fusion, five hidden layers are sequentially connected, ending in the final output layer.
Labels are extracted from browsers and incorporated into the data samples for training. The implementation of these labels generally adheres to standard definitions, such as IAB viewability and attention~\footnote{\raggedright\url{https://www.iab.com/wp-content/uploads/2024/08/IAB_Attention_Measurement_Explainer_August_2024.pdf}}.

\newaar{

The metrics selection for the model is at the publisher's discretion. We propose a framework using ad viewability as a key metric for effectiveness in reaching audiences and enhancing brand visibility. For instance, car ads aim for visibility, not immediate sales. Similarly, ads for car accident lawyers or government projects prioritize brand visibility where online conversion is not the aim. Furthermore, metrics for user interactions such as hovering, dwell time, and webpage scroll depth can also be supported by AdFL.

Additional metrics such as click-through rate, user attention, and post-click conversion rely on the publisher's capacity to monitor ad interactions beyond the initial session, which may involve accessing data from advertisers or third-party ad servers. In these scenarios, AdFL can address the issue in two ways: first, the publishers can transmit this data to the user’s storage and integrate it into the learning process in the browser. Alternatively, the publishers may design the model to partition user-related layers for browser-side execution, while managing external data processing on layers within the publisher’s server. In either approach, user data remains confined to their respective browsers.
}

\newaar{

\textbf{Aggregation: } Aggregation refers to the process by which local model updates are combined at the server-side. Each instance of aggregation constitutes a round and results in an updated global model. During the aggregation of local model parameters into the global model, each round involves participation from some or all users. The local model represents the user's model post-update, achieved by training with their individual data.

In AdFL, each client $i$ uploads its trained model parameters $\mathbf{\Theta}_i$, and the AdFL server aggregates the models using the FegAvg technique. FedAvg takes a weighted average of all models, using the size of each client's dataset as a weight to give more importance to models trained on larger datasets. In Equation~\ref{eqn:fedavg}, $\mathbf{\Theta}_i$ denotes the model parameters from the $i$-th client, $p_i$ represents the size of the dataset, and $N$ is the total number of clients. Clients can download the aggregated model parameters $\mathbf{\Theta}^{*}$ and use the model on their future data.

\begin{equation}
    \mathbf{\Theta}^{*} = \frac{\sum_{i=1}^{N}p_{i}\times\mathbf{\Theta}_i}{\sum_{i=1}^{N}p_{i}}
    \label{eqn:fedavg}    
\end{equation}

}

\section{Implementation}\label{sec:implementation}

In our implementation, we use libraries and APIs that are compatible with modern browsers. We implement AdFL using JavaScript for the client software and host the scripts at the server. Once the AdFL client is loaded to the user's browser, it leverages the browser's API such as the Event Listener and MutationObserver~\footnote{\raggedright\url{https://developer.mozilla.org/en-US/docs/Web/API/MutationObserver}, CC BY-SA} for monitoring data loading and element attribute or content modifications to facilitate data collection. 
The AdFL client utilizes IndexedDB~\footnote{\url{https://www.w3.org/TR/IndexedDB-2/}, W3C Document License} for storing data in the browser.
For model training and inference, the AdFL client employs TensorFlow.js
version 4.21 that is a downloadable script library and does not require permission to run. 

The AdFL client maintains data samples in a key-value format throughout the collection, processing, and storage processes. The components of the AdFL client interface with IndexedDB via the SM component. Algorithm~\ref{alg:ad_cycle} shows the pseudocode for running the AdFL client in the browser. 
The SM initializes the IndexedDB (Alg~\ref{alg:ad_cycle} Line 2), using a designated name and establishes the following data stores: `processedData', which retains data samples; `configuration', which contains model configurations and AdFL preferences; and `sessionData', which records temporal session data and the data of page requests from previous user visits to the publisher's site. These stores' names are used to load or store corresponding data as shown in line 6, 8, 15, and 20 of Algorithm~\ref{alg:ad_cycle}. The configuration, specifically maintained within 'config' as referenced in line 2, is utilized by all additional components. It is applied in line 3 for the setup of event listeners for ads, in lines 5 and 12 for the capturing of defined features, and in lines 7, 9, and 13 for both the computation and preprocessing of features.

\begin{algorithm}
\caption{AdFL's Ad Capture and Inference} 
\label{alg:ad_cycle}
\small
\begin{algorithmic}[1]
\STATE \textbf{Initialize:} Load to the browser AdFL Client and Ads 
\STATE $config \gets initSM()$ \COMMENT{//trigger to init other components}
\STATE $setAdMonitoring(config['adPlacesList'])$ \COMMENT{//such as 'adplacementTop'}

\STATE Capture Contextual data and setup the session data 
\STATE $userPageData \gets capture\_contextData( $ \\
    \hspace{9em} $ config['context'])$
\STATE $prevSession \gets load\_IndxdDB('sessionData')$
\STATE $currSession \gets computeSession( userPageData, $\\ 
    \hspace{6em} $prevSession, config['session'])$
\STATE $store\_IndxdDB(currSession, 'sessionData')$
\STATE $contxtFeatures \gets preprocess\_data( userPageData, $ \\ 
    \hspace{6em} $currSession, config['Features'])$ 
    
\WHILE{isTheSamePage()}
    \STATE \textbf{Capture and preprocess ad data:}
    \STATE $adData \gets capture\_adData(config['adData'])$
    \STATE $adFeatures \gets preprocess\_data(adData, $ \\ 
        \hspace{7em} $ config['Features'])$
    \STATE $newAdSample \gets concat(adFeatures, $ \\ 
        \hspace{7em} $ contxtFeatures)$
    \STATE $store\_IndxdDB(newAdSample, 'processedData')$
    \STATE \textbf{Perform inference:}
    \STATE async $invoker(tiggerModel(newAdSample))$
    \STATE \textbf{Capture Ad's Metrics:}
    \IF{metrics = MetricsReceived()}
        \STATE $update\_IndxdDb(metrics, newAdSample,$ \\
        \hspace{7em}$ 'processedData')$
    \ENDIF
    \STATE $wait\_for\_next\_ad()$
\ENDWHILE
\end{algorithmic}
\end{algorithm}

We conduct data validation and processing in the DPS's feature processing module, using predefined features and associated metadata for preprocessing. Due to limited access to complete datasets in FL, we use data-independent methods effective across users' datasets, such as hash functions, MinMax Scaling, and default values for missing data. These functionalities are encompassed within 'preprocess\_data' and are utilized in algorithm~\ref{alg:ad_cycle} on lines 9 and 13.

\newaa{
Publishers enable ads by allocating specific web elements within predefined locations on a webpage. These elements, such as \texttt{<div>} tags, are assigned unique identifiers (e.g., id=``adplacementTop") and shared with the ad server. The ad server incorporates these identifiers into its scripts to load ads into the specified \texttt{<div>} elements on the page. This ensures the ad server's script knows precisely where to load the ads. AdFL operates within this framework, utilizing the same list of identifiers to enable ad monitoring and set up ad-capturing mechanisms as in line 3 of algorithm~\ref{alg:ad_cycle}. AdFL does not work with Ad Blockers. It is designed as an alternative to Ad Blockers for publishers who would not show content unless the users turn the ad blockers off. Since this is a lose-lose proposition for publishers and users, AdFL protects user privacy and allows publishers to increase revenue.
}

The AdU monitors ads in each designated ad placement and throughout every refresh ad instance. 
Ads are inserted to the webpage and remain there for a predetermined duration known as the refresh rate, which is typically 15 seconds. Utilizing the native browser API, mutationObserver, the AdU monitors and records each instance of a new ad being loaded onto the page, as in line 11-15 of algorithm~\ref{alg:ad_cycle}. It evaluates and records various ad metrics, including viewability, clicks, and attention metrics, as in line 19-20. The data collected by the AdU includes the ad creative, the ad server query ID, the dimensions of the visitor's page, the dimensions of the ad, the proportion of the ad relative to the page dimensions, the placement ID, and the metadata concerning the attributes of the ad's nested containers.

Ads are rendered in the browser by initially loading the ad creative script into the data attributes of the container, as an intermediate stage prior to transferring the content to a secured iframe. AdU captures the ad's creative content during this interim stage. However, the size of the content vary and often provides limited information regarding the advertiser. This captured content primarily constitutes metadata about the ad, as it is filled with tracking and monitoring scripts. Capturing the actual ad is difficult since adTech agencies and advertisers employ unique methodologies in structuring and obfuscating the ad creative.

\newaar{
\textbf{Edge Cases:} 
If ad blockers are present, AdFL logs session data, minus the ads. This depends on the publisher's choice to serve users with ad blockers, either through a browser extension or natively. To ensure content access, publishers might require users to enable ads or subscribe. If ads are enabled, AdFL operates fully with them; otherwise, it runs minimal activities for future ad interactions.

For fault-tolerance, AdFL immediately stores data locally to handle abrupt session terminations, enabling resumption. AdFL stores model parameters at the server and keeps user data in the browser, preserving information once trained on, even if users or browsers do not retain data long-term. 

AdFL offers cross-device compatibility, leveraging libraries and standard storage mechanisms endorsed by major browsers for event triggers and local storage. AdFL employs first-party cookies instead of third-party cookies, which are restricted by certain browser environments. 
}
\section{Evaluation} \label{sec: experiment}

This section presents the experimental evaluation of AdFL and its proof-of-concept model. We conducted experiments to assess AdFL's feasibility in real-life by measuring its latency during feature preprocessing, training, and inference. Additionally, we measured the performance of the model to understand if AdFL can indeed support models for online ads with good accuracy. Furthermore, we conduct an evaluation of AdFL's performance with DP to understand the implications of such privacy safeguards. Finally, we evaluate the overhead associated with AdFL's communication costs.

\newaa{
Currently, there is no FL framework that works in the browser for online ads, as discussed in the Related Work. Therefore, we focused on micro-benchmark performance to understand AdFL behavior in real-time. Furthermore, we compared our proof-of-concept model with a Centralized Learning (CL) model. Centralized solutions, such as CL, are the best in terms of revenue because they allow the publishers/advertisers to target the ads, but they do not protect the user's privacy. AdFL on the other hand provides good privacy protection, while allowing providers to get a revenue close to what they would get with centralized solutions.
}
We have publicly released the AdFL model and a sample dataset in a repository~\footnote{\url{https://github.com/almrx/AdFL}}.

\subsection{AdFL Latency in the Browser}

\textbf{Experimental Settings.} We evaluate the latency of AdFL's operations using desktop and mobile browsers. The desktop setup uses the Edge browser on Windows 11 with a 64-bit 2.3 GHz CPU and 16 GB RAM (8 GB for the browser). The mobile uses Firefox on a Samsung S21 running Android with 8 GB RAM (4 GB for the browser). Using the model configured as shown in Figure~\ref{fig:model-mini}, we run feature preprocessing, training, and inference ten times each and report the average latency per operation on both platforms.

\begin{table}[h]
    \centering
        \begin{tabular}{lcc}
            \hline
            & \textbf{Desktop} & \textbf{Mobile}  \\ \hline
            Preprocess & 0.0395 $\pm$ 0.0105 & 0.0514 $\pm$ 0.0119 \\ 
            Training & 378.086 $\pm$ 16.916 & 468.986 $\pm$ 29.658  \\ 
            Inference & 2.072 $\pm$ 0.421 & 6.326 $\pm$ 0.744 \\ \hline
        \end{tabular}
    \caption{Average Time for Preprocessing Features (ms/sample), Training (ms/round) and Inference (ms/sample)}
    \label{tab:preprocessing_time}
\end{table}

\begin{table}[t]
    \centering
    \setlength{\tabcolsep}{1.5mm}
    \begin{tabular}{l|c|c}
        \hline
        \textbf{Memory Overhead} & \textbf{Desktop} & \textbf{Mobile} \\ \hline
        AdFL JS Memory Allocation 
        & 51.24 & 47.23 \\ 
        After loading dataset & 102.32 & 100.89 \\ 
        After training & 223.65 & 191.64 \\ 
        After training and inference & 232.28 & 198.52 \\ \hline
    \end{tabular}
    \caption{JS Memory Usage (MB) at Different Training Stages by Device}
    \label{tab:js_memory_usage}
\end{table}

\textbf{Results.} Table~\ref{tab:preprocessing_time}  shows that AdFL's performance is feasible in real-life.
As expected, we find superior performance on the desktop, due to its more powerful resources. Notably, the preprocessing task exhibits the minimal latency. This task was conducted on 65 features, with an average data sample length of 4,957 and a standard deviation of 248. 
Such preprocessing latency is considered feasible in a real-time scenarios, where sequential processing of data samples occurs concurrently with the loading of an ad's data, and any latency below 100ms is deemed acceptable. 

The inference latency is under 3ms for the desktop and under 7ms for the mobile, which encompasses both the feature preprocessing and the inference execution on the provided ads data. This demonstrates that 
AdFL does not disrupt the processes of loading online ads with significant overhead. Furthermore, after inference, AdFL can operate in under one second on an ad that generally remains active for 30 seconds, suggesting that sufficient time exists to execute publisher-defined invokers. These include predicting user engagement and attention metrics with the ad or reducing display of ads unlikely to be viewed due to being out of the viewport.

The training results show the execution latency per local training round within the browser. Specifically, the model undergoes 15 rounds of training employing user data, comprising 374 ads data samples.
The average duration necessary for training is less than 400 ms on desktop platforms and under half a second for mobile browsers. This is feasible for any practical scenario.

\newaa{Table~\ref{tab:js_memory_usage} presents the memory utilization required to run the AdFL script in the browser for both mobile and desktop devices and having the same setup as in Table~\ref{tab:preprocessing_time}. It demonstrates that AdFL utilizes approximately 50 MB out of the 4 GB and 1 GB of memory allocated for JavaScript on desktop and mobile devices, respectively. 
Furthermore, loading a dataset comprising 374 ads, which amounts to less than 1.5 MB, incurs an additional 50 MB of memory usage beyond that of AdFL. Subsequently, the model loading and execution of 15 training rounds necessitate approximately 100 MB of memory, with an inference execution requiring roughly an additional 10 MB. Collectively, these memory overheads accumulate to less than 200 MB for mobile devices and under 240 MB for desktop platforms, thereby rendering it feasible to operate on most contemporary devices.

}

\subsection{Model Evaluation}\label{subsec: model eval}

\textbf{Experimental Settings.}
In this experiment, the global FL model is initialized at the server and then each client engages in FL training using their data samples. To understand the model performance degradation due to FL, which provides improved privacy, we also run a centralized version of the model (CL). 

\begin{figure}[t!]
    \centering
    \begin{subfigure}[b]{0.48\linewidth}
        \centering
        \includegraphics[width=\linewidth]{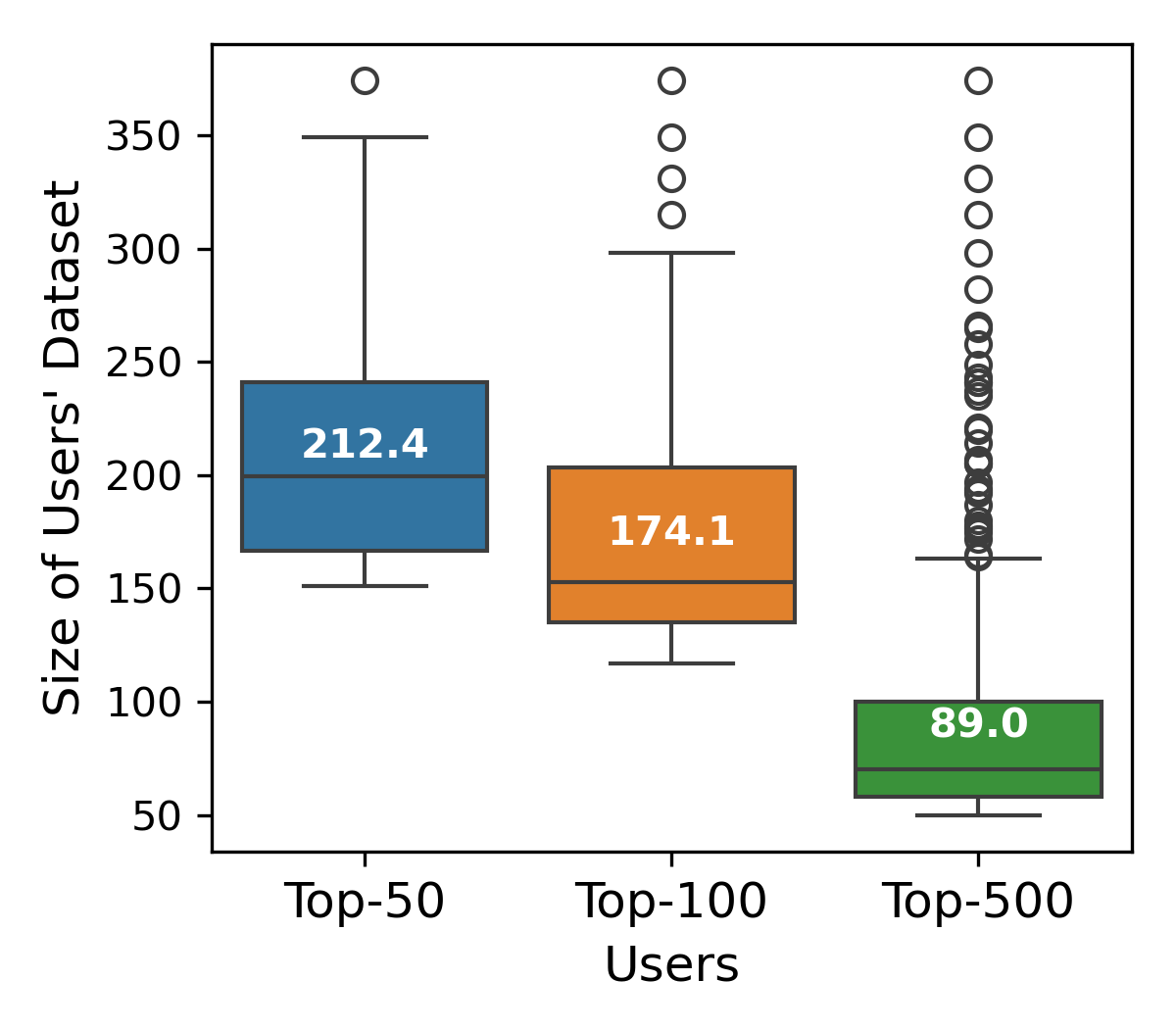}
        \caption{10-days}
        \label{fig:dataset_size_10days}
    \end{subfigure}
    \begin{subfigure}[b]{0.48\linewidth}
        \centering
        \includegraphics[width=\linewidth]{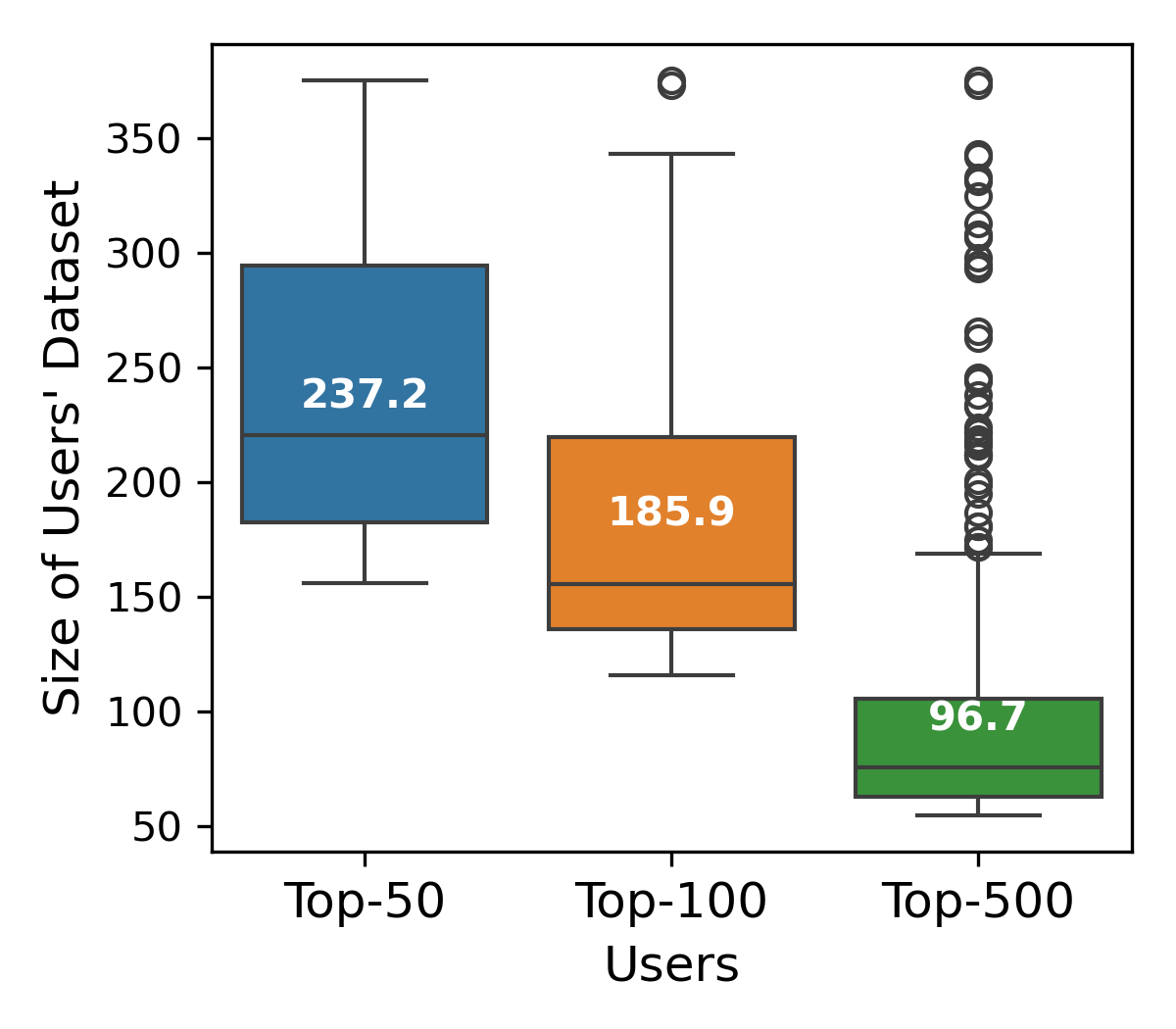}
        \caption{30-days}
        \label{fig:dataset_size_30days}
    \end{subfigure}
    \caption{Distribution of Data across the Two Datasets for Top-x Users (Mean Values inside the Boxes)}
    \label{fig:dataset_size}
\end{figure}

We collected two non-overlapping datasets of 10 days and  30 days on a publisher's website with weekly traffic of about 280,000 visitors, including 180,000 unique visitors, and 380,000 page views with 15,000 unique page views. 
The site offers static content without user login and does not store user data; thus, our datasets lacks user identifiers. Ads are served via the Prebid ad server, focusing on ad data and context. Having two datasets allows us to evaluate the model’s performance with different amounts of data and to assess whether the results are robust across different time periods.

Figure~\ref{fig:dataset_size} illustrates the range of the two dataset sizes categorized by user groups (users are ranked according to the size of their data). We segmented the data chronologically, and then split it into 80:10:10 for training, validation, and testing. AdFL configuration allows the server to define and transmit user selection criteria, such as data size, to the client. 
As expected, the figure shows that users accumulate more data over a 30-day period then over a 10-day period.

\begin{table}[t!]
    \centering
    \footnotesize
    \setlength{\tabcolsep}{1.5mm}
    \renewcommand{\arraystretch}{1.2}
    \begin{tabular}{lccc}
        \hline
        Averaged & \textbf{50 Users} & \textbf{100 Users} & \textbf{500 Users} \\
        \hline
        \textbf{Loss} & 0.4486$\pm$0.0016 & 0.5112$\pm$0.0042 & 0.5434$\pm$0.0020 \\
        \textbf{Accuracy} & 79.21$\pm$0.32 & 75.55$\pm$0.71 & 73.69$\pm$0.23 \\
        \textbf{AUC} & 87.36$\pm$0.12 & 82.96$\pm$0.24 & 80.78$\pm$0.13 \\
        \textbf{Rounds} & 123.3$\pm$17.5 & 212.2$\pm$49.2 & 252.0$\pm$62.1 \\
        \hline
    \end{tabular}
    \caption{FL Model Performance across Different User Settings (Input size = 27, Layers = 6, Parameters = 330K)}
    \label{tab:performance_metrics}
\end{table}

\textbf{Results.} Unless otherwise specified, the majority of the results are for the smaller dataset to demonstrate that the model performs well even with relatively low amounts of data.
Table~\ref{tab:performance_metrics} presents the performance metrics of the model for varying user counts of 50, 100, and 500. 
The results demonstrate that the model achieves good performance across several metrics. The slight decline in performance with the increase in the number of users is due to the selection of users for experiments based on their amount of data (e.g., top 50 users have more data than the next 50 users, etc.). In our experiment we collected data over a limited time period, but in real-life we expect each user to have more data than in these experiments. Therefore, the performance will be closer to the one achieved in the experiments for 50 users.

\begin{table}[t!]
    \centering
    \footnotesize
    \setlength{\tabcolsep}{2mm}  
    \renewcommand{\arraystretch}{1.2}  
    \begin{tabular}{lcccc}
        \hline
        \textbf{Model Settings} & \textbf{CL 10} & \textbf{CL 50} & \textbf{FL 10} & \textbf{FL 50} \\
        \hline
        \textbf{AUC} & 87.49 & 89.19 & 84.44 & 87.36 \\
        \hline
    \end{tabular}
    \caption{FL vs. CL Performance under Two User Settings}
    \label{tab:model_comparison}
\end{table}

\begin{figure}[tbp]
    \centering
    \includegraphics[width=0.75\linewidth]{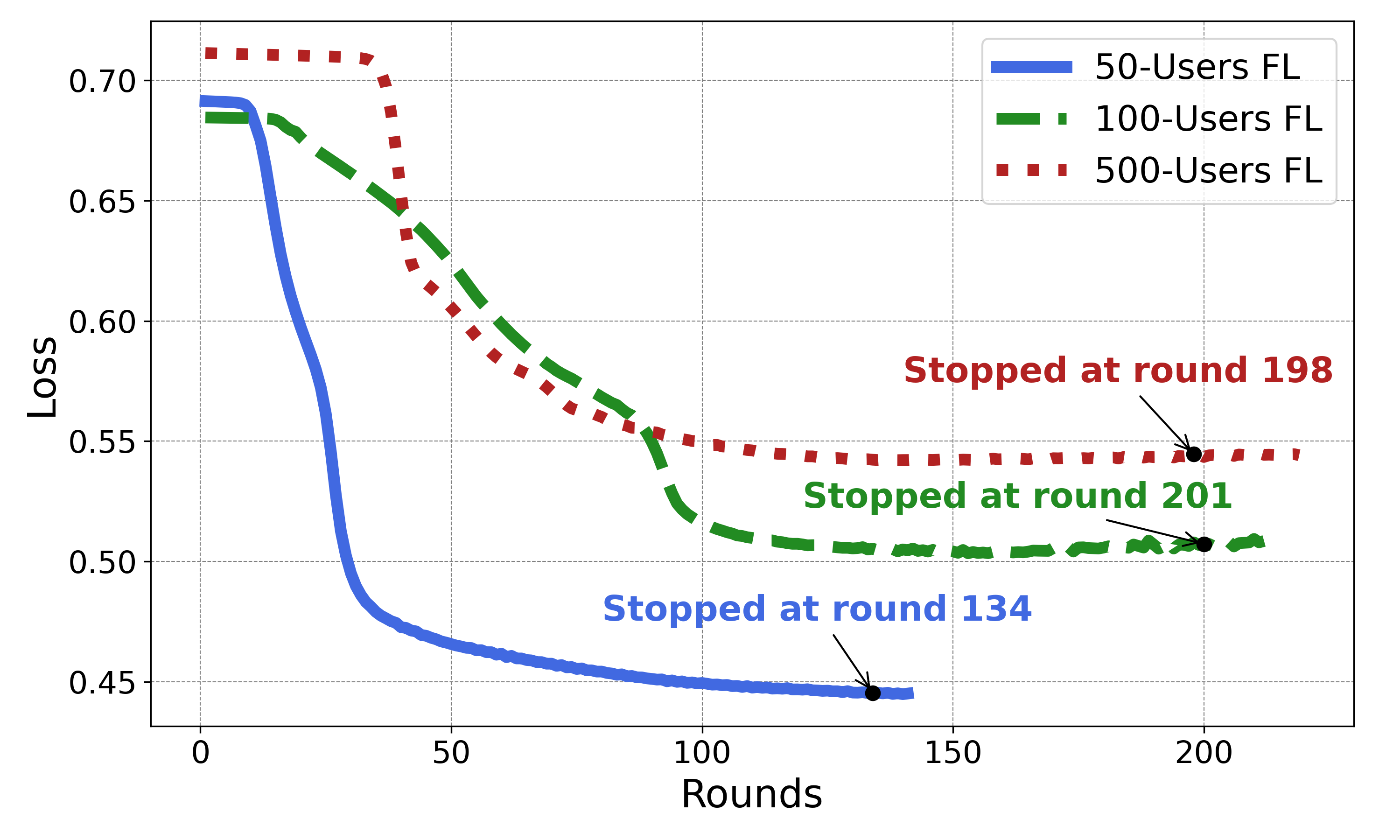}
    \caption{Convergence Rate for FL with Different \# of Users}
    \label{fig:convergence}
\end{figure}

Figure~\ref{fig:convergence} illustrates the convergence behavior of the model when trained with 50, 100, and 500 users. The model employs an early stopping mechanism, with patience parameters set at 7, 11, and 20 based on the validation loss, respectively. Notably, the dataset comprising 50 users represents a smaller user base with a well-balanced distribution of data. Conversely, the group with 500 users exhibits slower convergence. 
This category includes a substantial number of users with a minimal amount of data, beginning from 50 samples per user. Consequently, the overall data quality among participants improves the model's convergence efficiency. 

Table~\ref{tab:model_comparison} presents a comparison between the FL and the centralized (CL) version of the model to understand the FL degradation of performance vs. CL (i.e., the cost of privacy). The model parameters are the same as in Table~\ref{tab:performance_metrics}.   
The 10 users in the FL experiment are the users with the most data. CL10 and CL50 merge the users for FL10 and FL50 into one dataset.
AUC is recorded at the end of the training, which includes 100 rounds, with early stopping parameters set between 7 and 11 rounds of patience. The results demonstrate that the FL model on top of AdFL achieves good performance compared to the CL model, within a few percentage points.

\begin{table}[t!]
    \centering
    \footnotesize
    \begin{tabular}{cccccc}
        \hline
        Noise ($\epsilon$) & Rounds & Loss & AUC & Communication (MB) \\
        \hline
        0.1 & 117 & 0.5519 & 78.59 & 11797.41 \\
        0.5 & 180 & 0.5426 & 79.74 & 18177.13 \\
        1.0 & 263 & 0.5330 & 80.33 & 26582.15 \\
        \hline
    \end{tabular}
    \caption{FL Performance using Differential Privacy}
    \label{tab:differential_privacy}
\end{table}

\newaar{
While FL protects the local data of the users by design, the local model parameters can be used by the server to infer user-related data. A standard method to protect the local model parameters is to use Differential Privacy (DP). Table~\ref{tab:differential_privacy} illustrates the performance of the model involving 50 users under three DP levels, with the noise $\epsilon$ ranging from 0.1 to 1. This table provides a comparison using the applied noise, the number of global rounds, the global loss, the global performance (AUC), and the total cost of the data communication round expressed in megabytes (MB). An increase in the noise enhances the privacy of the model parameters and diminishes the capacity to infer individual user information. The model configuration is consistent with that outlined in Table \ref{tab:performance_metrics}. Overall, the performance of the model is still good, with a modest decrease in performance compared to the version without DP (i.e., the cost of stronger privacy). The experimental findings reveal that higher noise levels require additional rounds, consequently increasing the communication costs. The privacy cost when comparing the lower noise $\epsilon=0.1$ and the higher noise $\epsilon=1$ shows a 2.3-fold increase in the number of rounds required. Furthermore, the study demonstrates that an increased number of rounds contributes to enhanced model performance, as indicated by AUC and loss decline.
}

\begin{figure}[tbp]
    \centering
    \includegraphics[width=0.7\linewidth]{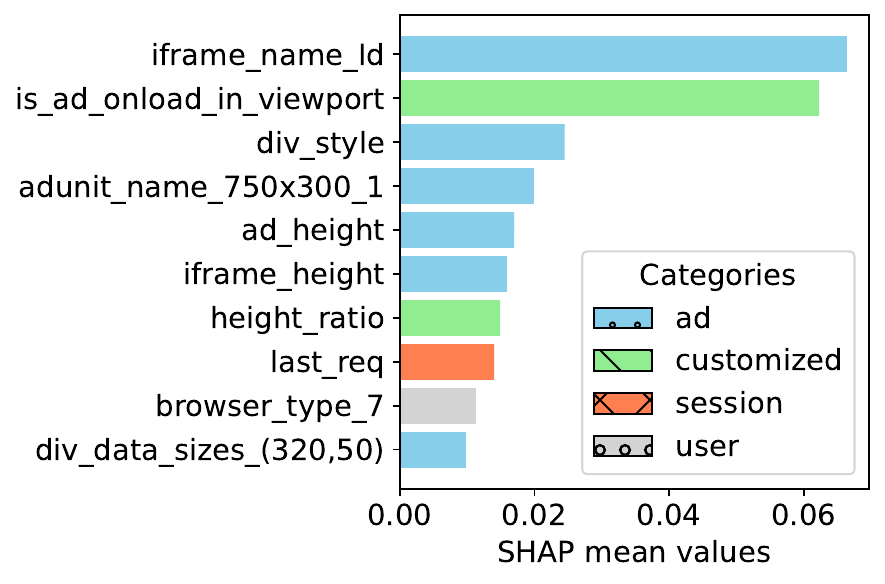}
    \caption{SHAP Feature Importance for 10-Day Dataset (Top 10)}
    \label{fig:feature shap view}
\end{figure}

To identify the key features for model efficacy, Figure~\ref{fig:feature shap view} employs the SHAP method to rank these features. The top ten features are categorized into ad, customized, session, user, and page categories (not shown). Customized features encapsulate specific metrics captured once during data collection, as opposed to labels, which necessitate a temporal condition. Features related to ads are the primary contributors to ad metrics, followed by customized, session, and user features. Conversely, page-related features are among the least consequential. Thus, AdFL's effort in collecting ad specific data is proved to be critical.

We performed a number of ablation studies to explore versions of the viewability-tasked model based on sizes and data inputs. 
Model versions included neural network layers, embedding layers, attention layers, and data segmentation. 
A BERT-based~\cite{antoun2020arabert} model was used for embeddings with textual inputs. 
Among configurations, the neural network layer excelled, with the fully connected model achieving an AUC of 87.55\% on the 10-days dataset, which include 50 users with dataset as refer to Figure~\ref{fig:dataset_size_10days}. Additionally, this lightwight model, referred to as miniModel and described in Section~\ref{subsec:model_design}, shows better performance than the same model with LSTM, Attention, and embedding layers. This model was thus used in the previous experiments because it is ideal for online advertising, as it requires less data and can run on resource-limited devices. The miniModel is designed to utilize a minimal set of features for input, and it employs the Adam optimizer without the imposition of a fixed learning rate. Experimental evaluations using learning rates of 1e-2, 1e-5, and 1e-6 revealed that the optimizer showed more effective convergence in the absence of a fixed learning rate within our experimental setup.

\begin{figure}[t!]
    \centering
    \includegraphics[width=0.95\linewidth]{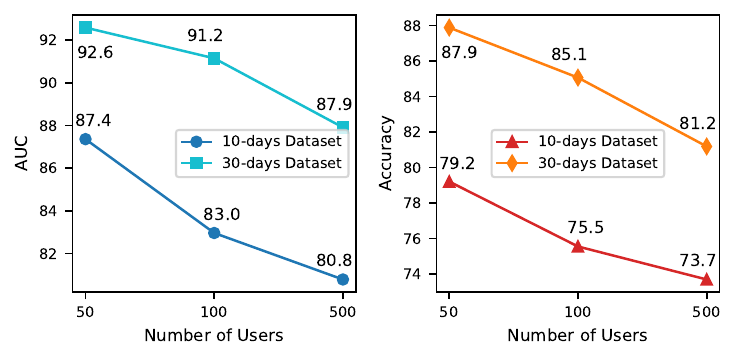}
    \caption{Model Performance 10-days vs. 30-days Datasets}
    \label{fig:datacomp}
\end{figure}

Figure~\ref{fig:datacomp} shows the comparison of model performance (AUC and accuracy) across the two datasets. We observe that the 30-days dataset leads to better performance, because each individual user has more data (as shown in Figure~\ref{fig:dataset_size}). Furthermore, since the datasets were collected at different times with no overlap, the results demonstrate that the model achieves robust performance over time.

\subsection{Discussion} \label{sec:discussion}

This work aims to enhance user privacy in the realm of online advertising, and it is aligned with existing policies and regulations, thereby respecting user rights and potentially yielding a positive impact on society. 

Regarding the limitations, the experimental data is confined to a particular publisher and its audience. Nonetheless, as long as enough data is available per user, we do not expect significant deviations in the results for other publishers, as the main model features are derived from ads rather than contextual information. 

\newaar{
In terms of communication cost, AdFL requires communication only during model download and updates, with model size impacting communication costs. These costs are part of the FL privacy trade-off and should be manageable with current browser resources. A model with 330,000 parameters ($\sim$1.5 MB) incurs about 3 MB per round trip. As shown in Table~\ref{tab:performance_metrics}, an FL model with 50 users, each contributing over 123 rounds, totals around 370 MB over multiple sessions. Typically, a user in a session participates in a dozen rounds due to time limits, resulting in an average data transfer of around 30MB, which is generally acceptable.
}

\section{Conclusion} \label{sec: conclusion}

This paper presented an end-to-end FL system for in-browser online ads, covering data collection, preprocessing, model training and inference. Our experimental results demonstrate that AdFL works well in real-time, which is important for the latency perceived by users. In addition, the proof-of-concept ad viewability model on top of AdFL achieves good performance, which makes is usable in applications that aim to better match ads with user interest and thus increase the publisher's revenue. Therefore, AdFL represents a promising practical solution toward balancing user privacy with publisher's revenue for targeted ads.

\section*{Acknowledgment}
This research was supported by the National Science Foundation (NSF) under Grants No. CNS 2237328 and DGE 2043104.

\bibliography{references}

\end{document}